\documentclass[preprint2]{aastex631}
\usepackage{booktabs}
\usepackage{changepage}
\usepackage[flushleft]{threeparttable}

\newcommand{\OII}{$\mathrm{[O\textsc{ii}]}$}
\newcommand{\OIII}{$\mathrm{[O\textsc{iii}]}$}
\newcommand{\Halpha}{H$\alpha$}
\newcommand{\Hbeta}{H$\beta$}
\newcommand{\Hdelta}{H$\delta$}

\newcommand{\totalsources}{$724$}

\newcommand{\robustmembers}{$40$}
\newcommand{\likelymembers}{$14$}
\newcommand{\totalHAEgrism}{$18$}
\newcommand{\numberquiescent}{$11$}

\newcommand{\quiescentfraction}{$62^{+20}_{-23}\%$}
\newcommand{\quiescentfractionlikely}{$50^{+19}_{-19}\%$}
\newcommand{\agnfraction}{$50^{+22}_{-22}\%$}
\newcommand{\tablefoot}[1]{\tablenotemark{\scriptsize {#1}}}

\defcitealias{shimakawa_mahalo_2018-1}{S18}

\begin{document}

\title{Revealing the quiescent galaxy population in the Spiderweb protocluster at $z=2.16$ with deep HST/WFC3 slitless spectroscopy}

\author[0000-0001-7713-0434]{Abdurrahman Naufal}
\affiliation{Department of Astronomical Science,
The Graduate University for Advanced Studies,
2-21-1 Osawa, Mitaka,
Tokyo 181-8588, Japan}
\affiliation{National Astronomical Observatory of Japan,
2-21-1 Osawa, Mitaka,
Tokyo 181-8588, Japan}

\author{Yusei Koyama}
\affiliation{Department of Astronomical Science,
The Graduate University for Advanced Studies,
2-21-1 Osawa, Mitaka,
Tokyo 181-8588, Japan}
\affiliation{Subaru Telescope,
National Astronomical Observatory of Japan,
650 North A’ohoku Place,
Hilo, HI 96720, USA}

\author{Chiara D'Eugenio}
\affiliation{Instituto de Astrofísica de Canarias (IAC), E-38205 La Laguna, Tenerife, Spain}
\affiliation{Universidad de La Laguna, Dpto. Astrofísica, E-38206 La Laguna, Tenerife, Spain}

\author{Helmut Dannerbauer}
\affiliation{Instituto de Astrofísica de Canarias (IAC), E-38205 La Laguna, Tenerife, Spain}
\affiliation{Universidad de La Laguna, Dpto. Astrofísica, E-38206 La Laguna, Tenerife, Spain}

\author{Rhythm Shimakawa}
\affiliation{Waseda Institute for Advanced Study (WIAS), Waseda University, 1-21-1, Nishi-Waseda, Shinjuku, Tokyo 169-0051, Japan}
\affiliation{Center for Data Science, Waseda University, 1-6-1, Nishi-Waseda, Shinjuku, Tokyo 169-0051, Japan}

\author{Jose Manuel Pérez-Martínez}
\affiliation{Instituto de Astrofísica de Canarias (IAC), E-38205 La Laguna, Tenerife, Spain}
\affiliation{Universidad de La Laguna, Dpto. Astrofísica, E-38206 La Laguna, Tenerife, Spain}

\author{Tadayuki Kodama}
\affiliation{Astronomical Institute, Tohoku University, 6-3, Aramaki, Aoba, Sendai, Miyagi 980-8578, Japan}

\author{Yuheng Zhang}
\affiliation{Purple Mountain Observatory, Chinese Academy of Sciences, 10 Yuanhua Road, Nanjing, 210023, China}
\affiliation{School of Astronomy and Space Science, University of Science and Technology of China, Hefei, Anhui 230026, China}
\affiliation{Instituto de Astrofísica de Canarias (IAC), E-38205 La Laguna, Tenerife, Spain}
\affiliation{Universidad de La Laguna, Dpto. Astrofísica, E-38206 La Laguna, Tenerife, Spain}

\author{Kazuki Daikuhara}
\affiliation{Astronomical Institute, Tohoku University, 6-3, Aramaki, Aoba, Sendai, Miyagi 980-8578, Japan}



\begin{abstract}

We report the HST WFC3 G141 grism slitless spectroscopy observation of the core region of the Spiderweb protocluster at $z=2.16$. We analyzed the spectra of all objects in a $\sim 2 \times 2 \text{ arcmin}^2$ field of view and identified \robustmembers{} protocluster members, recovering 19 previously identified \Halpha-emitters in addition to revealing 21 new members. The spectra allowed us to identify \numberquiescent{} galaxies with quiescent spectra. Three galaxies with quiescent spectra are possibly still star-forming according to SED fitting, indicating a possible left-over or dust-obscured star formation. We estimate a quiescent fraction of $\sim 50\%$ for $M_\star > 10^{11} M_\odot$. About half of the quiescent galaxies possibly host AGN, hinting at AGN's key role in quenching galaxies in the protocluster environment. These quiescent galaxies have relatively more compact and concentrated light profiles than the star-forming members, but they are not yet as bulge-dominated as local ellipticals. These results are consistent with previous studies that indicate the Spiderweb protocluster is in the maturing stage, with a red sequence that has begun forming.

\end{abstract}


\section{Introduction} \label{sec:intro}

Galaxy clusters are among the largest gravitationally bound structures in the Universe. It has been established that local galaxy clusters host a distinct galaxy population compared to the general field. Local galaxy clusters predominantly host massive elliptical, quiescent galaxies \citep[e.g., ][]{dressler_galaxy_1980, goto_morphology-density_2003}, contributing little to the local cosmic star-formation density \citep{kauffmann_environmental_2004, cybulski_voids_2014}. 

Cluster progenitors, i.e., protoclusters, are thought to play an important role in star-formation during the cosmic noon \citep{chiang_galaxy_2017}, forming and quenching massive galaxies at early epochs to become such quiescent environments today \citep[see ][]{overzier_realm_2016, alberts_clusters_2022}. Protoclusters have been observed at cosmic noon in variety of stages: some are found to be largely star-forming \citep[e.g., ][]{hayashi_starbursting_2012, dannerbauer_excess_2014, wang_discovery_2016, shimakawa_mahalo_2018, koyama_planck-selected_2021, polletta_spectroscopic_2021, daikuhara_star-formation_2024, perez-martinez_enhanced_2024} while others have been observed to be already forming a red sequence \citep[e.g., ][]{kodama_first_2007, willis_spectroscopic_2020, ito_cosmos2020_2023}. The high density of galaxies in protoclusters may boost interactions and mergers \citep{hine_enhanced_2016, liu_what_2023}, driving gas inflows to fuel starbursts and active galactic nuclei \citep[AGN; ][]{weston_incidence_2017, u_role_2022}. AGN feedback is thought to be the key mechanism in quenching massive galaxies and often needed by simulations to match the observed number densities of massive quiescent galaxies \citep[e.g., ][]{beckmann_cosmic_2017}. While AGN activity has been detected in high-redshift quiescent galaxies \citep{ito_cosmos2020_2022}, the relationship between protoclusters and AGN remains complicated to disentangle \citep[][see also \citealt{lovell_characterising_2018}]{alberts_star_2016, macuga_fraction_2019, shimakawa_new_2024}. In addition, major mergers may also transform star-forming disk galaxies into a more spheroidal structure, although AGN feedback is still needed to quench the star-formation \citep{lotz_galaxy_2008, lotz_effect_2010}.

The protocluster PKS 1138-262 is originally identified as an overdensity of Lyman-$\alpha$ emitters around a radio galaxy, dubbed the Spiderweb Galaxy \citep{miley_spiderweb_2006}, at redshift $z=2.16$ \citep{kurk_search_2000, pentericci_search_2000}. Since then, Spiderweb protocluster has become one of the best-studied protoclusters with various surveys unveiling diverse galaxy populations such as \Halpha-emitters \citep{kurk_search_2004-1, koyama_massive_2013, shimakawa_mahalo_2018-1}, X-ray-emitters \citep{pentericci_search_2000, tozzi_700_2022}, submillimeter galaxies and CO-emitters \citep{dannerbauer_excess_2014, dannerbauer_implications_2017, emonts_giant_2018, tadaki_environmental_2019, jin_coalas_2021}, and photometrically red galaxies \citep{kurk_search_2004-1, kodama_first_2007, tanaka_formation_2013}. The Spiderweb protocluster is one of the protoclusters targeted by MApping H-Alpha and Lines of Oxygen with Subaru (MAHALO-Subaru) narrowband survey \citep{kodama_mahalo-subaru_2013} and its extension MAHALO Deep Cluster Survey \citep{shimakawa_mahalo_2018, daikuhara_star-formation_2024}.

In this paper, we report the results of Hubble Space Telescope (HST) Wide Field Camera 3 (WFC3) G141 grism slitless spectroscopy of a $4 \text{ arcmin}^2$ region at the core of Spiderweb protocluster. Such observation is aimed to find and spectroscopically confirm the population of galaxies in the protocluster core in an unbiased manner, i.e., without the need of pre-selection of objects, down to the near-infrared continuum flux limit. This observation allows us to identify quiescent galaxies in the protocluster.
This paper is organized as follows: We describe the dataset and data reduction in Section \ref{sec:data}. In Section \ref{sec:membership}, we explain the selection criteria for protocluster members. We present the identified members from slitless spectroscopy in Section \ref{sec:results}. We then discuss the nature of quiescent galaxies in this protocluster in \ref{sec:discussion} and summarize our findings in Section \ref{sec:conclusion}. When needed, we assume a Planck15 cosmology \citep{planck_collaboration_planck_2020} with parameters of $H_0 = 67.7 \text{ km} \text{ s}^{-1} \text{ Mpc}^{-1}$, $\Omega_\lambda = 0.69$, and $\Omega_m=0.31$. We use AB magnitude system \citep{oke_secondary_1983} and \citet{chabrier_galactic_2003} (IMF) initial mass function throughout the paper.

\section{Dataset and data reduction} \label{sec:data}

\subsection{HST WFC3 G141}

\begin{figure*}
    \centering
    \includegraphics[scale=0.85]{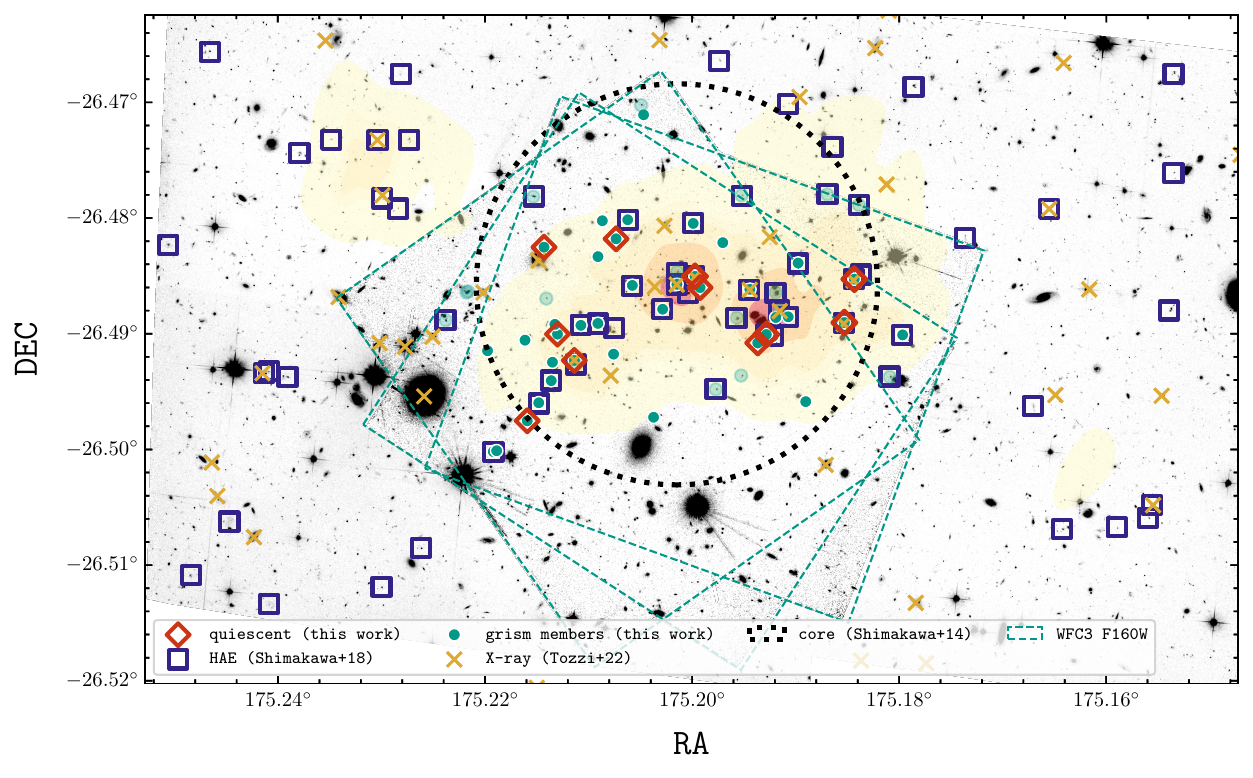}
    \caption{The spatial distribution of grism-identified members (teal circles) plotted on the $I_{814}$+$H_{160}$ image mosaic. Blue squares and yellow crosses denote HAEs \citepalias{shimakawa_mahalo_2018-1} and X-ray sources \citep{tozzi_700_2022}, respectively. Spectroscopically identified quiescent galaxies are marked by red diamonds. Grism-identified members follow the same spatial distribution as HAEs as shown by the yellow contours. The black dotted circle has a radius of $R_{200}$ from \citet{shimakawa_identification_2014}. Teal dashed squares show the HST WFC3 imaging in three different position angles.}
    \label{fig:spatial-distribution}
\end{figure*}

The Spiderweb protocluster was observed by the Hubble Space Telescope (HST) Wide Field Camera 3 (WFC3) during Cycle 30 in slitless spectroscopy mode using G141 grism covering a $\sim 2 \times 2 \; \text{arcmin}^2$ area of the core region \citep[Proposal ID 17117,][]{koyama_complete_2022}. G141 grism has a spectral range of $10750$\AA{} to $17000$\AA{},  with a nominal spectral resolution of $R\sim 130$, covering the $4000$ \AA{} break, \OII, \Hdelta, \Hbeta, and \OIII{} lines for galaxies at $z \sim 2.16$.

The observation was performed in 7 orbits in three different orientation angles to mitigate the effect of spectral contamination. For each orientation, the spectroscopic observation is accompanied by a direct imaging in $H_{160}$ filter. The total exposures are of $4.3$ hours and $40$ minutes in the slitless spectroscopy and direct imaging modes, respectively, in the deepest region of the mosaic.

The region has a wealth of multiwavelength photometry available taken by ground- and space-based telescopes (see Table \ref{tab:photometry}). The observation footprint covers $32$ narrowband-selected HAE members of the protocluster \citep[][hereafter S18]{shimakawa_mahalo_2018-1}, $18$ X-ray-detected sources \citep{tozzi_700_2022}, and $15$ CO-emitters \citep{jin_coalas_2021}, as shown in Figure \ref{fig:spatial-distribution}. For the rest of this paper, HAE refers to narrowband-selected HAEs identified by \citetalias{shimakawa_mahalo_2018-1}.

\begin{deluxetable}{llll}
    \tablecaption{Photometric bands used for SED fitting with CIGALE.} \label{tab:photometry}
    \scriptsize
    \tablehead{\colhead{Instrument} & \colhead{Filter} & \colhead{Exptime} & \colhead{Reference}}
    \startdata
    Chandra/    &           &           & \\
    ACIS-S      & hard  &     & \citet{tozzi_700_2022} \\
    ...         & soft &     & \\
    HST/        &           &           & \\
    ACS         & $g_{814}$     & 20670     & \citet{miley_spiderweb_2006} \\
    ...         & $I_{814}$     & 23004     & ... \\
    WFC3        & $H_{160}$     & 2490      & This work \\
    Subaru/     &           &           & \\
    S-cam       & $B$       & 6300    & \citetalias{shimakawa_mahalo_2018-1} \\
    ...         & $z'$      & 4500    & \citet{koyama_massive_2013} \\
    MOIRCS      & $J$       & 9060    & \citet{kodama_first_2007} \\
    ...         & $K_s$     & 3300    & \citet{kodama_first_2007} \\  
    ...         & $\mathrm{NB}_{2071}$  & 11160   & \citet{koyama_massive_2013}, \citetalias{shimakawa_mahalo_2018-1} \\
    VLT/        &           &           & \\
    HAWKI       & $Y$       & 26880     & \citet{dannerbauer_implications_2017} \\
    ...         & $H$       & 14830     & ... \\
    Spitzer/     &           &           & \\
    IRAC        & $3.6\mu$m & 3000      & \citet{seymour_massive_2007} \\
    ...         & $4.5\mu$m & 3000      & ...
    \enddata
\end{deluxetable}


\subsection{Data reduction}
We process the G141 and $H_{160}$ data using the \textsc{Python} package Grism redshift and line analysis \textsc{Grizli} v1.8.14 \citep{brammer_grizli_2019} which provides a full end-to-end processing of space-based slitless spectroscopy data. In short, \textsc{Grizli} is responsible for three steps: preprocessing, contamination modeling, and redshift fitting. 

\textsc{Grizli} preprocesses all the raw WFC3 exposures, i.e., performing astrometric alignment, flat-fielding, and sky subtraction. \textsc{Grizli} also creates a mosaicked $H_{160}$ direct image from which the source detection is performed through SEP \citep{barbary_sep_2016}, a Python implementation of SExtractor \citep{bertin_sextractor_1996}. 

For each source detected in the mosaic with $H_{160}$ magnitude $H_{160} < 26$, \textsc{Grizli} generates a 2D model spectrum for each exposure based on its measured flux density and iteratively refines it by fitting a third-order polynomial profile. \textsc{Grizli} subtracts all the contaminating model spectra to extract the spectrum of an object of interest. 

We extract the spectra of all sources in the mosaicked direct image field-of-view with $H_{160}$ magnitude $< 26$ (\totalsources{} sources) and determine the redshift using \textsc{Grizli}. To determine the redshift, \textsc{Grizli} fits the spectrum by a set of Flexible Stellar Population Synthesis continuum templates \citep[FSPS; ][]{conroy_propagation_2010} projected to 2D grid based on the object's morphology in the direct image. Simultaneously, in this iteration, \textsc{Grizli} also fits line complex templates, i.e., $\mathrm{H\alpha+N\textsc{ii}+S\textsc{ii}+S\textsc{iii}+He}$, \OIII$+$\Hbeta, and \OII+Ne with fixed flux ratios to solve for degeneracies in the redshift. We allow \textsc{Grizli} to fit the redshift within $0.05 < z < 5.00$ with a step of $0.01$. At each step, \textsc{Grizli} fits the templates with non-negative least squares, calculates the $\chi^2$ of the fit, and derives the probability density. Around the maximum peak of the probability density distribution, \textsc{Grizli} then performs the fitting again with a smaller step of $0.001$ using continuum templates and emission line templates with unrestricted line ratios. The result of the second iteration gives the best-fitting template and a posterior distribution for the redshift, as well as line fluxes. Hereafter, we quote the redshift at maximum posterior.

\section{Analysis}

\subsection{Membership selection} \label{sec:membership}

\begin{deluxetable*}{cccccccccccc}
\centering
\tablecaption{Robust members of the Spiderweb protocluster identified by HST G141}\label{tab:catalog_robust}
\tabletypesize{\scriptsize}
\tablehead{\colhead{ID} & \colhead{ID S18} & \colhead{$H_{160}$} & \colhead{$z_\mathrm{grism}$} & \colhead{$z_\mathrm{lit}$} & \colhead{$P_\mathrm{cl}$} & \colhead{$D_n4000$} & \colhead{quiescence} &  \colhead{AGN} & \colhead{$\log M_\star / M_\odot$} & \colhead{$r_\mathrm{half-light}$ (kpc)}}
\startdata
210 & 16 & 23.05 & $2.166^{+0.003}_{-0.006}$ & 2.157 & 0.99 & $0.81 \pm 0.12$ & -- & -- & $9.962^{+0.145}_{-0.220}$ & $2.72 \pm 0.05$ \\
211 & -- & 24.28 & $2.158^{+0.003}_{-0.005}$ & -- & 0.98 & $1.34 \pm 0.69$ & -- & -- & $9.219^{+0.135}_{-0.197}$ & $1.98 \pm 0.07$ \\
258 & -- & 23.00 & $2.155^{+0.003}_{-0.004}$ & -- & 1.00 & $1.40 \pm 0.16$ & Q? & -- & $10.292^{+0.070}_{-0.083}$ & $1.68 \pm 0.04$ \\
266 & -- & 24.49 & $2.191^{+0.012}_{-0.248}$ & -- & 0.57 & $1.18 \pm 0.46$ & -- & -- & $9.490^{+0.118}_{-0.162}$ & $2.27 \pm 0.20$ \\
296 & 21 & 23.33 & $2.179^{+0.005}_{-0.013}$ & 2.158 & 0.81 & $0.85 \pm 0.10$ & -- & -- & $10.068^{+0.086}_{-0.107}$ & $3.43 \pm 0.08$ \\
297 & -- & 23.85 & $2.177^{+0.003}_{-0.004}$ & -- & 0.89 & $1.70 \pm 0.43$ & -- & -- & $9.928^{+0.102}_{-0.133}$ & $1.57 \pm 0.05$ \\
335 & 30 & 21.93 & $2.170^{+0.014}_{-0.022}$ & 2.153 & 0.73 & $0.77 \pm 0.12$ & -- & -- & $11.170^{+0.149}_{-0.229}$ & $6.88 \pm 0.12$ \\
364 & -- & 23.68 & $2.155^{+0.003}_{-0.003}$ & -- & 0.98 & $0.90 \pm 0.21$ & -- & -- & $9.882^{+0.139}_{-0.206}$ & $3.13 \pm 0.10$ \\
369 & -- & 22.53 & $2.175^{+0.003}_{-0.005}$ & -- & 0.99 & $1.33 \pm 0.11$ & Q+AGN & X & $10.945^{+0.064}_{-0.075}$ & $1.54 \pm 0.02$ \\
386 & -- & 24.18 & $2.156^{+0.001}_{-0.001}$ & -- & 0.92 & $0.49 \pm 0.19$ & -- & -- & $8.719^{+0.143}_{-0.215}$ & $1.97 \pm 0.07$ \\
392 & -- & 24.04 & $2.149^{+0.012}_{-0.004}$ & -- & 0.92 & $1.01 \pm 0.21$ & -- & -- & $9.522^{+0.115}_{-0.157}$ & $2.97 \pm 0.10$ \\
411 & -- & 23.96 & $2.147^{+0.001}_{-0.001}$ & -- & 1.02 & $0.91 \pm 0.37$ & -- & M & $10.205^{+0.163}_{-0.265}$ & $17.83 \pm 6.43$ \\
412 & -- & 21.86 & $2.162^{+0.004}_{-0.002}$ & -- & 1.00 & $1.21 \pm 0.07$ & Q? & -- & $11.330^{+0.070}_{-0.083}$ & $4.32 \pm 0.14$ \\
418 & -- & 24.45 & $2.159^{+0.002}_{-0.002}$ & -- & 0.99 & $0.75 \pm 0.28$ & -- & -- & $8.236^{+0.110}_{-0.147}$ & $8.93 \pm 2.04$ \\
429 & 35 & 22.10 & $2.152^{+0.001}_{-0.001}$ & 2.155 & 1.00 & $0.98 \pm 0.05$ & -- & M & $10.353^{+0.117}_{-0.160}$ & $1.94 \pm 0.02$ \\
432 & 32 & 23.39 & $2.163^{+0.005}_{-0.070}$ & -- & 0.67 & $0.76 \pm 0.20$ & -- & -- & $9.537^{+0.111}_{-0.149}$ & $4.03 \pm 0.18$ \\
440 & -- & 23.13 & $2.165^{+0.009}_{-0.023}$ & -- & 0.72 & $1.15 \pm 0.26$ & Q & -- & $10.802^{+0.068}_{-0.081}$ & $2.31 \pm 0.06$ \\
443 & 39 & 22.30 & $2.156^{+0.008}_{-0.008}$ & -- & 0.96 & $1.25 \pm 0.21$ & Q? & -- & $10.709^{+0.143}_{-0.215}$ & $4.02 \pm 0.08$ \\
459 & 33 & 23.75 & $2.156^{+0.005}_{-0.003}$ & -- & 0.99 & $0.85 \pm 0.22$ & -- & -- & $9.623^{+0.122}_{-0.171}$ & $2.00 \pm 0.05$ \\
461 & -- & 24.98 & $2.191^{+0.003}_{-0.002}$ & -- & 1.00 & $0.19 \pm 0.49$ & -- & -- & $8.935^{+0.112}_{-0.151}$ & $5.94 \pm 0.26$ \\
465 & 38 & 23.59 & $2.158^{+0.001}_{-0.001}$ & 2.155 & 1.00 & $1.06 \pm 0.28$ & -- & -- & $9.337^{+0.105}_{-0.139}$ & $2.72 \pm 0.10$ \\
467 & 40 & 22.49 & $2.166^{+0.004}_{-0.005}$ & 2.162 & 0.98 & $1.33 \pm 0.17$ & Q+AGN & X & $10.999^{+0.076}_{-0.092}$ & $1.71 \pm 0.03$ \\
479 & -- & 24.44 & $2.165^{+0.006}_{-0.010}$ & -- & 0.82 & $0.71 \pm 0.31$ & -- & -- & $9.983^{+0.193}_{-0.355}$ & $2.52 \pm 0.12$ \\
482 & 41 & 23.41 & $2.156^{+0.006}_{-0.006}$ & -- & 0.95 & $1.01 \pm 0.15$ & -- & -- & $9.869^{+0.133}_{-0.192}$ & $2.13 \pm 0.05$ \\
503 & 42 & 23.43 & $2.166^{+0.003}_{-0.004}$ & -- & 0.99 & $0.87 \pm 0.18$ & -- & M & $10.301^{+0.178}_{-0.306}$ & $4.18 \pm 0.12$ \\
551 & 48 & 22.73 & $2.160^{+0.004}_{-0.004}$ & 2.166 & 0.99 & $0.77 \pm 0.11$ & -- & X & $11.013^{+0.083}_{-0.103}$ & $4.72 \pm 0.30$ \\
557 & -- & 22.72 & $2.154^{+0.006}_{-0.017}$ & -- & 0.85 & $1.67 \pm 0.27$ & Q+AGN? & M & $11.061^{+0.057}_{-0.065}$ & $4.75 \pm 0.20$ \\
558 & 49 & 23.71 & $2.166^{+0.001}_{-0.001}$ & 2.166 & 0.98 & $0.58 \pm 0.16$ & -- & -- & $9.670^{+0.172}_{-0.289}$ & $2.93 \pm 0.06$ \\
569 & 55 & 21.66 & $2.158^{+0.001}_{-0.001}$ & 2.169 & 0.99 & $1.43 \pm 0.12$ & Q+AGN & X & $11.351^{+0.063}_{-0.073}$ & $3.04 \pm 0.06$ \\
577 & 73 & 19.15 & $2.163^{+0.000}_{-0.000}$ & 2.156\tablefoot{a} & 0.67 & $1.01 \pm 0.01$ & -- & X+M & $12.435^{+0.117}_{-0.161}$ & $2.34 \pm 0.02$ \\
588 & 58 & 22.30 & $2.144^{+0.001}_{-0.001}$ & 2.157 & 1.00 & $1.17 \pm 0.15$ & Q+AGN & X+M & $10.833^{+0.123}_{-0.173}$ & $3.08 \pm 0.06$ \\
610 & 57 & 23.85 & $2.153^{+0.001}_{-0.001}$ & 2.152 & 1.02 & $1.43 \pm 0.69$ & -- & -- & $9.538^{+0.160}_{-0.257}$ & $2.60 \pm 0.06$ \\
624 & -- & 24.47 & $2.149^{+2.709}_{-0.004}$ & -- & 0.68 & $0.98 \pm 0.38$ & -- & -- & $9.258^{+0.124}_{-0.173}$ & $5.12 \pm 0.78$ \\
642 & -- & 21.39 & $2.147^{+0.004}_{-0.002}$ & -- & 1.00 & $1.87 \pm 0.17$ & Q & -- & $11.312^{+0.053}_{-0.060}$ & $4.58 \pm 0.08$ \\
650 & -- & 24.33 & $2.160^{+0.002}_{-0.002}$ & -- & 1.00 & $1.17 \pm 0.52$ & -- & -- & $9.331^{+0.139}_{-0.206}$ & $6.59 \pm 1.36$ \\
654 & -- & 22.36 & $2.151^{+0.004}_{-0.008}$ & -- & 1.00 & $1.35 \pm 0.15$ & Q & -- & $11.068^{+0.067}_{-0.079}$ & $2.46 \pm 0.04$ \\
682 & 65 & 23.54 & $2.183^{+0.003}_{-0.005}$ & 2.163 & 0.99 & $0.82 \pm 0.29$ & -- & -- & $10.152^{+0.182}_{-0.319}$ & $3.26 \pm 0.07$ \\
686 & -- & 24.26 & $2.150^{+0.003}_{-0.003}$ & -- & 0.98 & $2.42 \pm 1.65$ & -- & -- & $9.734^{+0.124}_{-0.173}$ & $1.55 \pm 0.08$ \\
691 & 64 & 23.70 & $2.173^{+0.010}_{-0.004}$ & -- & 0.87 & $0.54 \pm 0.25$ & -- & -- & $9.919^{+0.165}_{-0.270}$ & $4.54 \pm 0.28$ \\
770 & -- & 23.81 & $2.113^{+0.002}_{-0.001}$ & -- & 1.00 & $0.80 \pm 0.23$ & -- & -- & $9.835^{+0.097}_{-0.125}$ & $2.44 \pm 0.10$
\enddata
\tablecomments{$z_\mathrm{lit}$ list spectroscopic redshifts from \citet{perez-martinez_signs_2023} and \citet{shimakawa_mahalo_2018-1} (prioritizing the former for objects present in both).  $P_\mathrm{cl}$ refers to Equation \ref{eq:prob}. $D_n4000$ is the strength of the $4000 \text{\AA}$ break measured by \textsc{Grizli}. In the quiescence column, we mark quiescent galaxies with the `Q' appended by `+AGN' if they show AGN signatures. `Q?' indicates quiescent galaxies with high SFRs derived from SED-fitting.  In the AGN column, `X' refers to X-ray detected AGN \citep{tozzi_700_2022} and `M' refers to AGN candidates based on MEx diagram (see \ref{sec:mex}), which should be considered tentative. Stellar masses are estimated from SED fitting with CIGALE (see \ref{sec:sed}). Sérsic half-light radii are measured by GALFIT.} \tablenotetext{a}{Spiderweb Galaxy. Spectroscopic redshift taken from \citet{liu_new_2002}.}
\end{deluxetable*}

Ground-based observations have spectroscopically confirmed narrow-band selected HAEs having the redshifts of $2.14 < z < 2.17$ \citep{shimakawa_identification_2014, perez-martinez_signs_2023}. However, \citet{jin_coalas_2021} found that the overdensity in the distribution of CO-emitters is more extended, $2.10 < z < 2.21$, suggesting a possible super-structure. Following this possibility, we select galaxies with redshift within $2.10 < z < 2.21$ as determined by \textsc{Grizli}. We employ a magnitude cut $H_{160} < 25$ for the membership selection, as fainter sources have more noisy continuum and the even fainter contaminating sources might not be sufficiently modeled.

\textsc{Grizli} determines the redshift of an object as the redshift where the posterior probability density is at maximum. However, the spectrum quality is reflected in the posterior redshift distribution: a spectrum with strong emission lines and/or spectral break will have a strong single peak in the posterior redshift distribution while a spectrum with only weak features or very noisy will have flatter distribution with spikes. A spectrum with a single strong emission line may also have several strong peaks in the posterior distribution, i.e., showing a redshift degeneracy. To mitigate this, we calculate the probability of the redshift within the chosen range by integrating the probability density $p(z)$:

\begin{equation} \label{eq:prob}
    P_\mathrm{cl} = P(2.10 < z < 2.21) = \int_{2.10}^{2.21} p(z) \,dz.
\end{equation}
We consider objects with $P(2.10 < z < 2.21) > 0.5$ as potential members of the protocluster instead of relying on the posterior maxima, as similarly employed by \citet{willis_spectroscopic_2020}. 

The HAEs in the HST field-of-view can serve as a benchmark to check the membership selection. Narrowband HAE selection ensures that these objects have a strong emission line at $\lambda \sim 20710$ \AA{}, corresponding to \Halpha{} line at $z=2.155 \pm 0.20$ \citep[see ][]{shimakawa_mahalo_2018}. Out of $32$ HAEs in the field of view, \totalHAEgrism{} have grism redshifts fulfilling the membership criteria. 

We inspected the 14 HAEs not selected by the membership criteria. Among this sample, one HAE (ID 536) has its spectra in all orientations contaminated by the very extended \OIII{} emission of the Spiderweb Galaxy itself, adding extended emission lines at $\lambda \sim 14500$ \AA{} which causes \textsc{Grizli} to identify it as \Hbeta$+$\OIII{} line at $z = 1.874$. ID 338 has its spectrum cut off for $\lambda > 13500$ \AA{}. We discard these two HAEs and consider only 30 HAEs in the field of view.

Two other HAEs (ID 296 and 595) seem to have unsubtracted contamination when we check the 2D spectra by eye. For these two objects, the spectrum in one orientation has a strong emission line that does not exist in two other orientations, which may be caused by contamination by zeroth order dispersion of another object masquerading as an emission line because it lies on the beam of the object of interest in that particular orientation. When discarding the particular orientation and redoing the fitting with \textsc{Grizli}, it determines that both objects have redshifts $z=2.178^{+0.013}_{-0.005}$ and $2.190^{+0.013}_{-0.005}$, within the redshift selection range, although the latter has $P_\mathrm{cl} < 0.5$. There are also two other HAEs (ID 362 and 545) within the redshift selection range but have $P_\mathrm{cl} < 0.5$. Assuming all narrowband-selected HAEs are true members of the protocluster and if we include these four HAEs recovered above, the recovery rate is around $73 \%{} (22/30)$, after accounting for the cases above. 
This is slightly lower than the rough estimate of the recovery rate ($\sim$83\%) assuming all HAEs show strong \OIII{}-emission lines with \OIII{}/\Halpha{}$ =1$. The recovery rate we mentioned above would be reasonable, given the fact that \OIII{} emissions of HAEs vary in strength \citep{suzuki_o_2016}, in addition to contamination affecting the redshift determination.

We also performed such visual inspection on the 2D and 1D spectra of the non-HAE galaxies fulfilling the criteria. We examined if there is a visible contamination in the spectrum or if the spectrum is cut off due to being located at the edge of the image. We discard two galaxies in this step and do not consider them as potential members. 

In addition, we ran the same pipeline with different source detection parameters (see Appendix \ref{sec:likely}). Objects selected by both the default configuration and the alternate configurations are considered `robust', while those selected by only one are considered as `likely'. Narrowband-selected HAEs with visually discernible \Hbeta+\OIII-like emission lines but not recovered by grism redshift selection are also considered as `likely' members.

After these careful checking steps, we consider \robustmembers{} galaxies as robust members of the Spiderweb protocluster identified by HST grism. We consider \likelymembers{} galaxies as likely members of the protocluster (see Appendix \ref{sec:likely}). 

\subsection{Physical parameters} \label{sec:sed}

We derive physical parameters, e.g., stellar mass, of the selected members by SED fitting using Code Investigating GALaxy Emission \citep[CIGALE, ][]{boquien_cigale_2019}. As listed in Table \ref{tab:photometry}, we use photometric bands from optical to NIR for the SED fitting. For X-ray sources detected by \citet{tozzi_700_2022}, we also include X-ray fluxes in the SED fitting.

We fit stellar population synthesis templates of \citet{bruzual_stellar_2003}, assuming \citet{chabrier_galactic_2003} IMF and a delayed tau-model star-formation history, to the photometric data with redshifts values fixed to those derived by \textsc{Grizli}. For likely members, we fix the redshift value to $z=2.16$. We fix the metallicity to $Z=0.004 \;(0.2 Z_\odot)$. We set the maximum age of the main stellar population to be the age of the Universe at the redshift, $\sim 3$ Gyr, and e-folding times between $0.1$ and $6$ Gyr. We include nebular templates with ionization parameter $\log U=-2.0$ and adopt \citet{calzetti_dust_2000} dust attenuation law with $E(B-V)=0.1-1.0$. We include AGN component \citep{stalevski_dust_2016, fritz_revisiting_2006} in the fitting only for X-ray sources, with AGN fraction range of $0.01-0.9$.

To examine the stellar-mass--size relation (Section \ref{sec:size}), we ran \textsc{GALFIT} \citep{peng_detailed_2010} on the $H_{160}$ image to obtain the Sérsic index and half-light radius of grism-selected members. For each galaxy, we fit a single Sérsic profile with index constrained within $0.2 < n < 6.0$ and mask all other detected sources surrounding the target. We use the empirical PSF from \citet{anderson_empirical_2016} in the fitting. In this analysis, we assume the $H_{160}$ light profile is dominated by the stellar continuum. The uncertainties in half-light radius and Sérsic index are estimated from bootstrapping analysis by perturbing the image with random Gaussian sky noise.

\section{Results} \label{sec:results}

\subsection{Spatial and redshift distribution} \label{sec:distribution}

\begin{figure}
        \centering
        \includegraphics[width=\columnwidth]{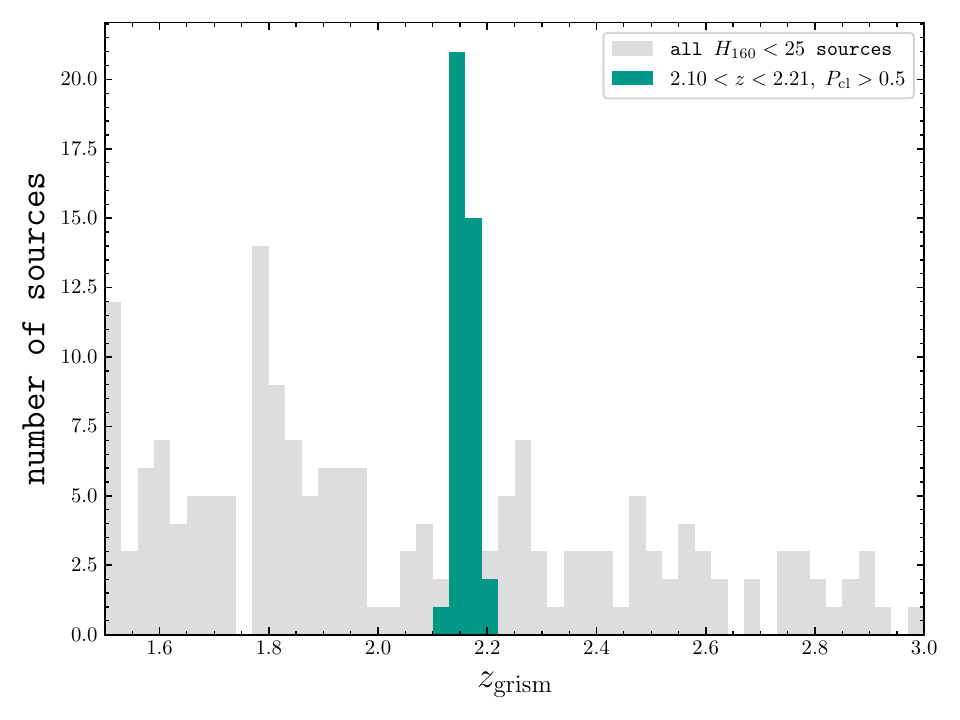}
        \caption{Distribution of redshifts determined by \textsc{Grizli}. We select galaxies with redshifts $2.10 < z < 2.21$ The distribution of selected robust members is shown in teal histogram while non-member sources with $H_{160} < 25$ is shown in gray.}
    \label{fig:redshift}
\end{figure}
We show the spatial distribution of robust members and likely members in Figure \ref{fig:spatial-distribution}, as well as sources detected in several previous surveys \citep{shimakawa_identification_2014, shimakawa_mahalo_2018-1, jin_coalas_2021, tozzi_700_2022}. Table \ref{tab:catalog_robust} lists the robust members selected by grism spectroscopy. Similar to HAEs, new grism-identified members seem to be distributed in a filamentary structure extending eastward and westward from the Spiderweb Galaxy \citep{croft_filamentary_2005, koyama_massive_2013}, while being sparse on the southern side, supporting their classification as protocluster members.

The redshift distribution of sources determined by \textsc{Grizli} is shown in Figure \ref{fig:redshift}. As described in Section \ref{sec:membership}, the redshift range for membership selection is based on the possible super-protocluster or line-of-sight filamentary structure found by \citet{jin_coalas_2021}. While this redshift range is wider than the narrowband redshift coverage, we find that 31 out of \robustmembers{} are within the narrowband coverage, i.e., we do not find significant overdensity in the extended redshift distribution due to the smaller field of view of the HST observation, as most CO-emitters identified by \citet{jin_coalas_2021} lie outside the protocluster core. Nonetheless, we confirm that 4 CO-emitters are robust members with $z_\mathrm{grism} = 2.159-2.169$ in addition to 2 which are likely members with $z_\mathrm{grism} = 2.189$ and $2.218$.

Based on grism membership classification, the projected number density of this protocluster for $H_{160} < 25$ is 16 (6) times higher than the average density of general field in the narrowband (extended) redshift slice based on to 3D-HST catalog \citep{momcheva_3d-hst_2016}.

\subsection{\OIII-emitters and active galactic nuclei} \label{sec:mex}

\begin{figure}
    \centering
    \includegraphics[width=\columnwidth]{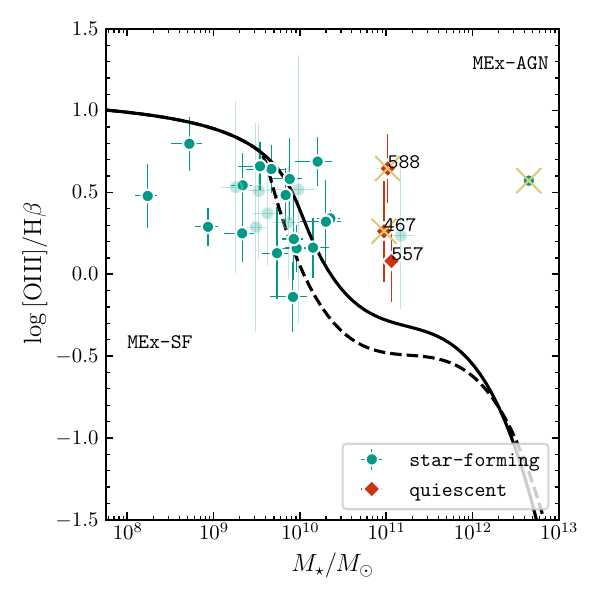}
    \caption{Mass-excitation diagram for \OIII-emitters in our sample. Solid line demarcates star-forming region and AGN region, while the region between solid line and dashed line is considered as intermediate \citep{juneau_active_2014}. Teal circles represent grism-identified star-forming members with emission line signal-to-noise ratio $>2$. Red diamonds represent the spectroscopically selected quiescent grism members with \OIII{} emission lines. Transparent data points have line $S/N>1$. X-ray-detected HAEs are expectedly located in the AGN region along with some new possible AGN identified in our sample.}
    \label{fig:MEx-diagram}
\end{figure}

Among \robustmembers{} robust members, we identified 8 \OIII-emitters which are not in the HAE catalog of \citetalias{shimakawa_mahalo_2018-1}. These \OIII-emitters are among the less massive protocluster members, with stellar masses of $9 \lesssim \log M_\star/M_\odot \lesssim 10$. Two of these objects have grism redshifts outside the range of $\mathrm{NB}_{2071}$ filter used to select HAE members in \citetalias{shimakawa_mahalo_2018-1}. The remaining six \OIII-emitters are not selected by \citetalias{shimakawa_mahalo_2018-1} as they do not fulfill the $K_s$ detection limit or the $Bz'K$ selection (see Section 2.3 of \citetalias{shimakawa_mahalo_2018-1}). Our analysis of grism data has shown that they are protocluster members at $z\sim 2.16$, implying they indeed might be HAEs.


We check the nature of \OIII-emitters and HAEs by plotting them in the Mass-Excitation diagram \citep[MEx; ][]{juneau_new_2011, juneau_active_2014} in Figure \ref{fig:MEx-diagram}. Here, we only show galaxies with emission line $S/N$ $>1$. This method can identify X-ray-faint AGN with stellar mass, \OIII{}, and \Hbeta{} line fluxes information. Since grism spectrum cannot resolve \OIII$\lambda5007$ and \OIII$\lambda4959$ separately, we correct the \OIII{} flux considering the flux ratio \OIII$\lambda5007 /$\OIII$\lambda4959 = 2.98$ \citep{dimitrijevic_flux_2007} to obtain \OIII$\lambda5007$ flux. X-ray detected HAEs are located in the MEx-AGN region in addition to four non-X-ray members. Combining X-ray-detected robust members and MEx-identified sources, we identified 10 AGN candidates in the core of the protocluster. We should note that the SF-AGN segregration with MEx is less clear at $z\sim2$ \citep{juneau_active_2014, coil_mosdef_2015}, so these new MEx-AGN should only be considered tentative.

\subsection{Quiescent galaxies} \label{sec:quiescent}

\begin{figure*}[t]
    \centering
    \epsscale{1.1}
    \plottwo{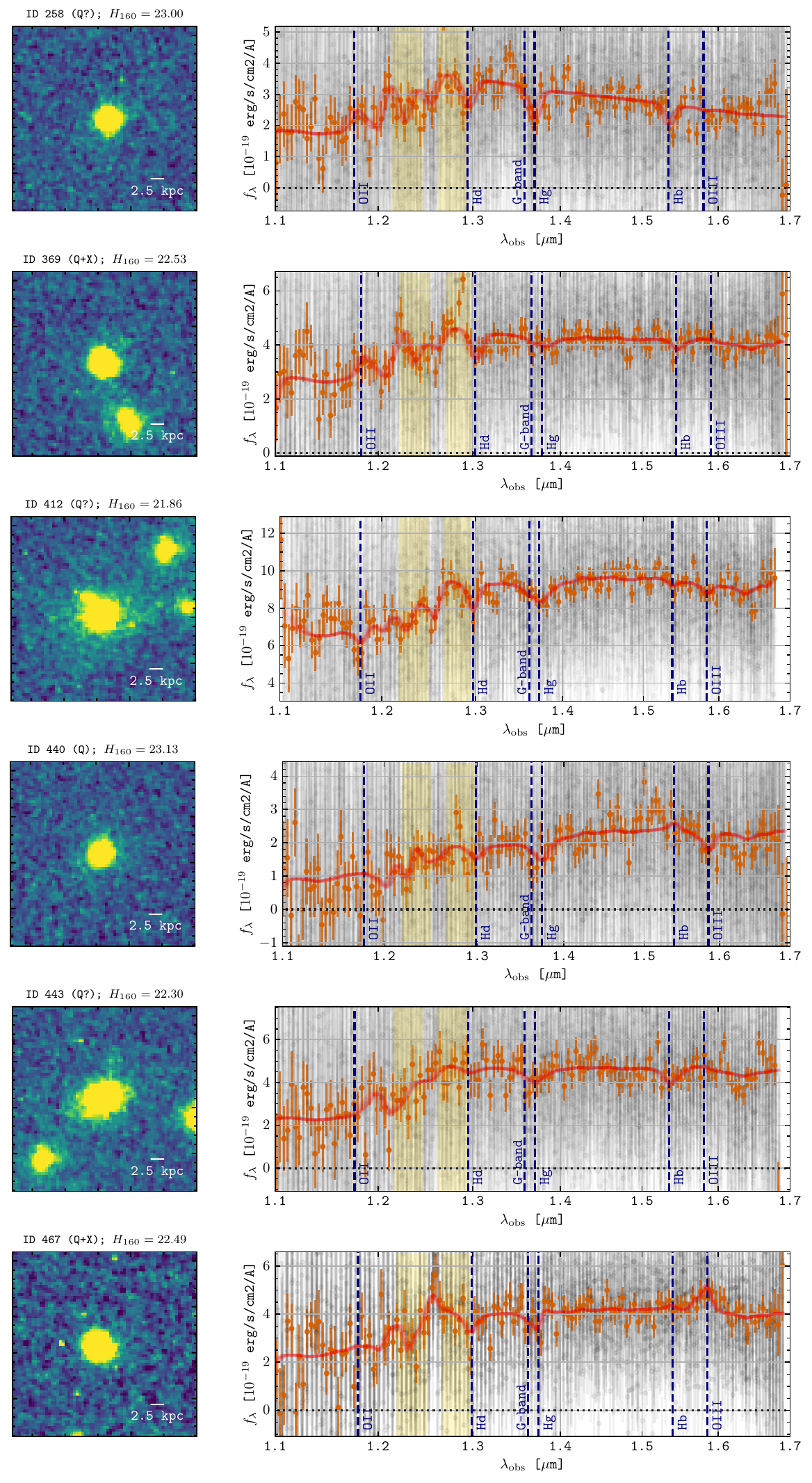}{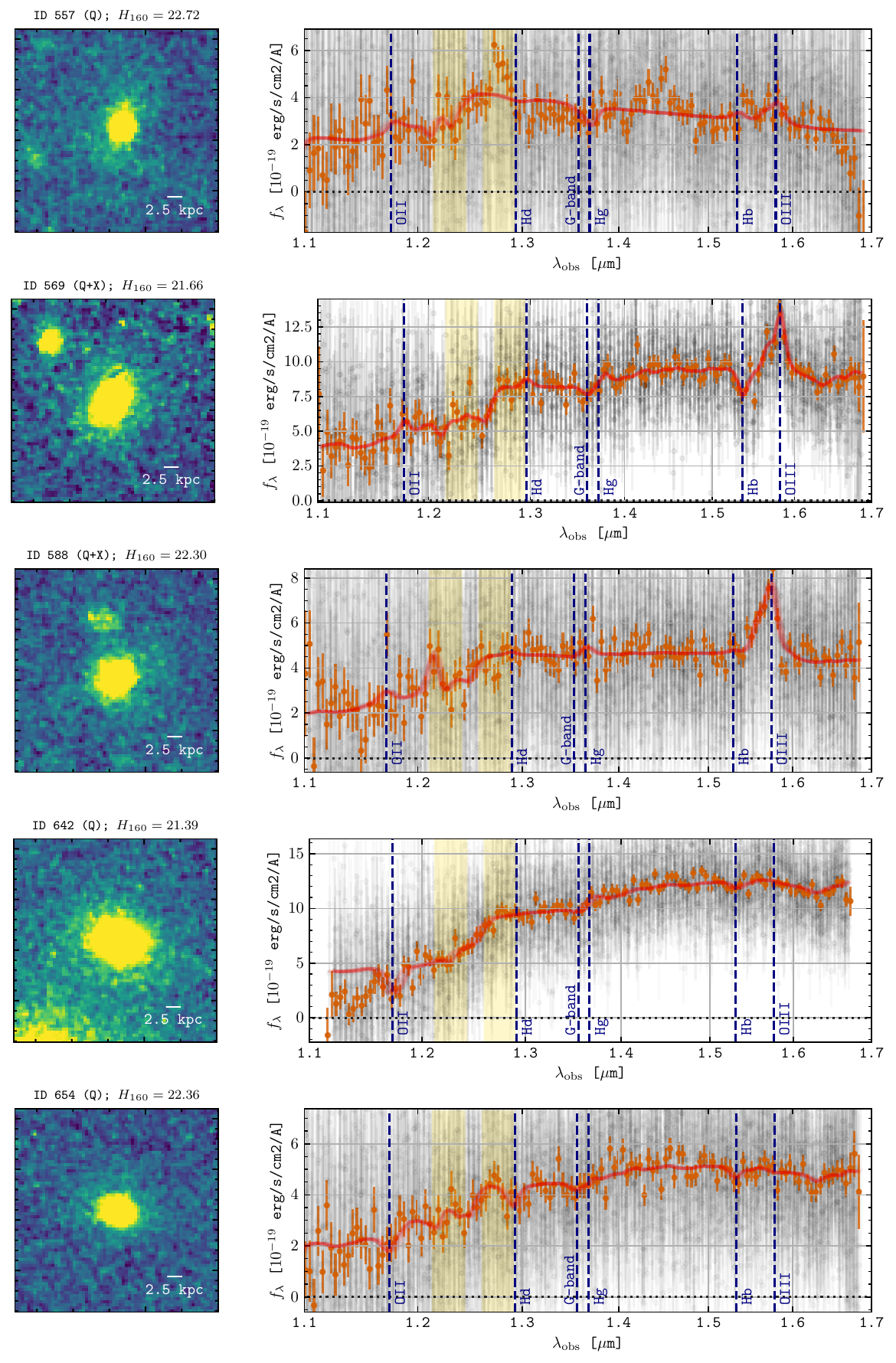}

    \caption{Here we show the spectra of the quiescent candidates in Spiderweb protocluster identified by HST WFC3 G141 grism observation. For each object, the left panel shows a $5 \times 5 \text{ arcsec}^2$ cutout of the direct image. The right panel shows the 1D spectrum of the stacked (orange) and individual (gray) exposures, with red line denoting the best-fitting FSPS templates. Blue vertical lines mark some salient lines in the wavelength range, regardless of their detection in the particular spectrum. Yellow-shaded regions mark the regions for calculating $D_n4000$ \citep{balogh_differential_1999}. }
    
    \label{fig:showcase1}

\end{figure*}

The main goal of the HST grism observation of this protocluster region is to search for quiescent galaxies residing in this overdense environment. To identify quiescent candidates, we search for member galaxies with $4000 \; \text{\AA}$ break strength $D_n4000 > 1.1$ (as defined by \citet{balogh_differential_1999}) and has no emission lines, except when the object is detected in X-ray \citep{tozzi_700_2022} or is located in MEx-AGN region in the mass-excitation diagram (see Section \ref{sec:mex}), to entertain the possibility of quenched galaxies hosting AGN. Based on this criteria, we identify \numberquiescent{} quiescent galaxies in the protocluster, including 4 HAE+AGN quiescent galaxies. We show the spectra and the direct image of quiescent members in Figure \ref{fig:showcase1}. Quiescent members are shown as red diamonds in Figure \ref{fig:spatial-distribution}, demonstrating they are distributed in a similar manner to other members galaxies.

All the selected quiescent galaxies have stellar mass of $\log M_\star / M_\odot \gtrsim 10.2$. Our approach is biased against quiescent galaxies with lower stellar mass due to the continuum limit to detect $4000 \; \text{\AA}$ breaks. On the other hand, star-forming galaxies with lower stellar mass can still have detectable emission lines, such as the \OIII-emitters we discussed in Section \ref{sec:mex}, thus reducing the dependency on the stellar continuum.

On average, the quiescent sample has $D_n4000 \approx 1.37$. The brightest quiescent candidate in our catalog, ID 642, was identified by \citet{tanaka_formation_2013} as a potential quiescent galaxy based on Subaru/MOIRCS data. The grism spectrum shows a $D_n4000 = 1.87 \pm 0.17$ without strong Balmer absorption lines. These suggest that this galaxy has relatively old stellar population and has ceased its star formation activity for a longer time than other members, although the lack of absorption lines may also be caused by morphological smoothing along the spectral axis. The other quiescent galaxies are likely to be more recently quenched, as their spectra show Balmer absorption lines and exhibit weaker $D_n4000$. 

The X-ray-detected HAE ID 569 is discussed by \citet{shimakawa_new_2024} as a quiescent galaxy with an AGN based on Keck/MOSFIRE J-band spectroscopy. Our grism spectrum support this further as it exhibits a prominent $4000 \; \text{\AA}$ break and \Hbeta{} apparently in absorption, albeit with a strong \OIII{} emission. We suspect that ID 467 and 588, two other X-ray-detected HAEs, are also candidates of galaxies quenched or being quenched by AGN as their grism spectra are similar to ID 569. One of the newly identified MEx-AGN candidate, ID 557, may also be a quiescent+AGN candidate as it has $D_n4000 = 1.67 \pm 0.27$, fulfilling our criteria. 

\begin{figure}
    \centering
    \includegraphics[width=\columnwidth]{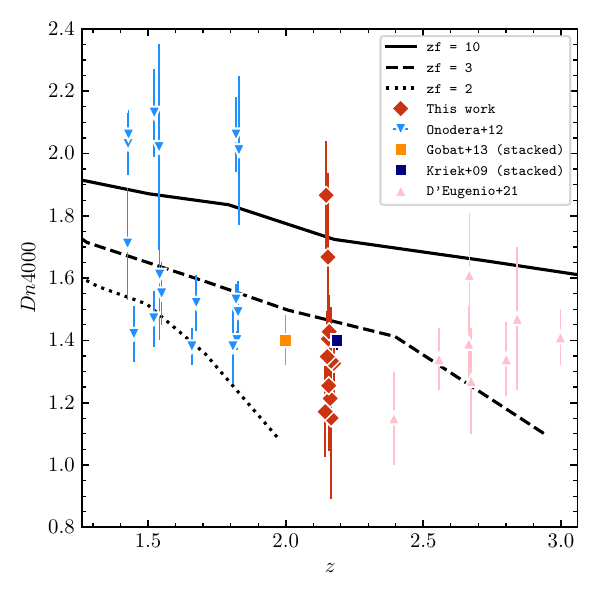}
    \caption{$D_n4000$ as a function of redshift, following Figure 13 of \citet{gobat_wfc3_2013}. The solid, dashed, and dotted lines show the evolution of $D_n4000$ of single stellar population synthesis (assuming solar metallicity) formed in $z_f = 10, 3, \text{ and } 2$, respectively. We also show results from \citet{kriek_direct_2006}, \citet{onodera_deep_2012}, \citet{gobat_wfc3_2013}, and \citet{deugenio_hst_2021}. The quiescent galaxies in Spiderweb protocluster have $D_n4000 \approx 1.37$, similar to a single stellar population with $z_f = 3$, implying that they are recently quenched.}
    \label{fig:gobat}
\end{figure}

In Figure \ref{fig:gobat}, we compile from several studies the $4000 \; \text{\AA}$ break of high-redshift quiescent galaxies, to put our results in context of the evolution of quiescent galaxies. As a visual aid, we add the evolution of $D_n4000$ for single stellar population models with different formation redshift $z_f$. The $D_n4000$ of our quiescent sample are roughly consistent with the model with $z_f=3$, implying relatively recent quenching, in agreement with other observations of high-redshift quiescent galaxies \citep[e.g., ][]{kriek_direct_2006, deugenio_typical_2020, deugenio_hst_2021}. 

\section{Discussion} \label{sec:discussion}

\subsection{Quiescence of the newly identified protocluster members} \label{sec:quiescence}

\begin{figure}
    \centering
    \includegraphics[width=\columnwidth]{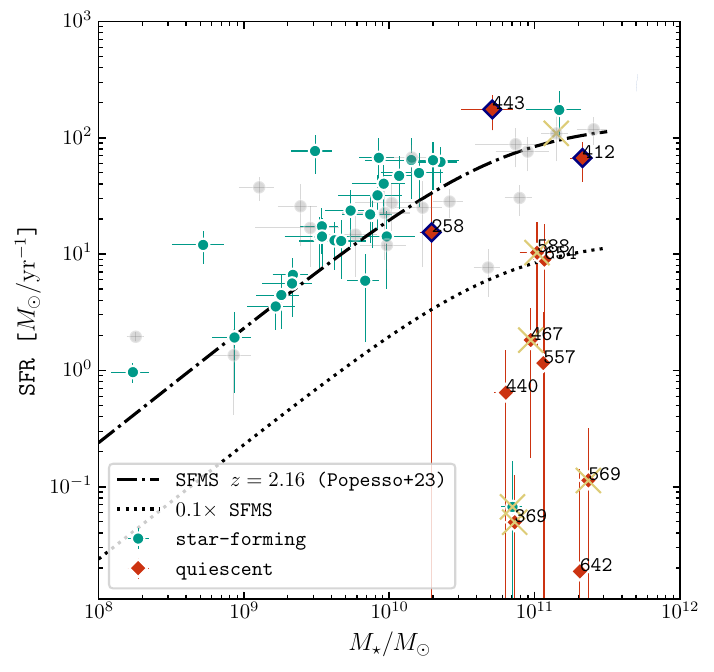}
    \caption{Stellar mass--star-formation rate diagram showing estimates derived from CIGALE. Spectrum-selected quiescent galaxies are shown as red diamonds, with blue borders indicating they are located near SFMS. Other robust members are shown in teal circles and likely members in semi-transparent gray. The Spiderweb Galaxy is outside the range of this plot ($M_\star > 10^{12} M_\odot$) and is omitted for brevity.}
    \label{fig:sfr}
\end{figure}

In Section \ref{sec:quiescent}, we identified \numberquiescent{} galaxies as quiescent based on their spectra. As a further check, we derived the star-formation rate and stellar mass by fitting the SED with CIGALE (see \ref{sec:sed}). We show the stellar mass--star-formation rate in Figure \ref{fig:sfr}. The majority of selected quiescent candidates lie below the star-formation main sequence (SFMS) at this redshift \citep{popesso_main_2023}, agreeing with the quiescence identification by the spectrum. However, there are three spectroscopically selected quiescent galaxies that lie on SFMS according to CIGALE: IDs 443, 412, and 258.

HAE ID 443, satisfies our quiescence criteria with apparent Balmer absorption and no emission lines, but is not identified as AGN. It is possible that this galaxy is a dusty galaxy instead of a quiescent one, similar to the dusty e(a) galaxies in \citet{poggianti_optical_2000} with no emission lines in rest-frame $3000-5000 \text{ \AA}$ while emitting \Halpha. 

\begin{figure}
    \centering
    \includegraphics[width=\columnwidth]{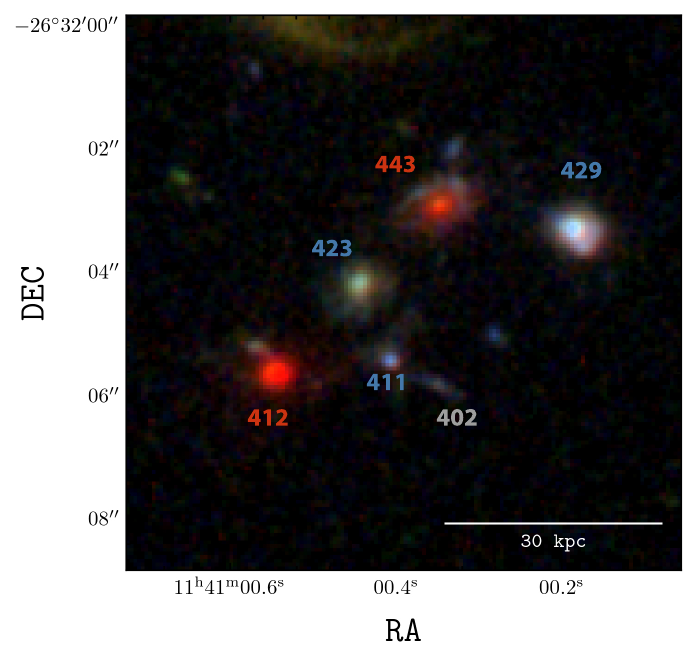}
    \caption{An $9\times9 \text{ arcsec}^2$ cutout of the dense complex including the quiescent candidates ID 412 and 443, a blue HAE ID 429, an \OIII{}-emitter ID 411, and a likely member ID 423. The composite color image is based on WFC3 $H_{160}$ for red, ACS $I_{814}$ for green, and ACS $g_{475}$ for blue, with arbitrary scaling to enhance the bluer part for clarity. The ACS images are downsampled to match the pixel scale of WFC3.}
    \label{fig:complex}
\end{figure}

We also inspected the multi-color images of this galaxy in Figure \ref{fig:complex}. While the $H_{160}$ image of this object shows a smooth morphology, ACS $g_{475}$ and $I_{814}$ images \citep[rest-frame UV, ][]{miley_spiderweb_2006} reveals that this object consists of several bright star-forming clumps. Combined together, the object exhibits a strong red core surrounded by star-forming clumps. Since \textsc{Grizli} downweights the outskirts part of the 2D spectrum during extraction \citep{horne_optimal_1986}, the spectrum is dominated by the core of the galaxy. Instead of a dusty galaxy, this system could also be an extreme example of inside-out quenching galaxy or a quiescent galaxy rejuvenated by a merger.

As also shown in Figure \ref{fig:complex}, ID 412 exhibits an asymmetric disturbance akin to a tidal tail, which is dominated in rest-frame UV light in the composite RGB. In $g_{475}$, only the tail is detected, while the main part is not. It is possible that the main galaxy is already quenched, while star-formation is happening in the tail component triggered by a merger. This disturbance is unresolved in ground-based and lower resolution imaging. Since the photometric data is based on $\mathrm{NB}_{2071}$ detection, it includes the star-forming part, causing the high estimated SFR.

The local environment of these two quiescent candidates is also very interesting: within $6 \text{ arcsec}^2$ radius ($\approx 50$ kpc), there are two other robust members; the HAE ID 429 and the \OIII-emitter ID 411; as well as ID 423 at $z=2.229$ ($P_\mathrm{cl} = 0.26$) and ID 402 which is very faint ($H_{160} = 25.5$) but with a strong emission line at $\lambda \sim 15800 \; \text{\AA}$ (\OIII{} line at $z=2.16$). This group of galaxies may eventually merge into one with $M_\star \sim 3 \times 10^{11} M_\odot$.

ID 258 is the least massive of our spectra-selected sample. While the spectrum shows Balmer absorption lines, it is relatively bright in $g_{475}$ and $I_{814}$, unlike other quiescent sample. This galaxy is likely a post-starburst galaxy, with some leftover young stars dominating the rest-frame UV. We should note that the $\mathrm{SFR}$s quoted here is estimated by CIGALE by Bayesian analysis from the probability density function, but CIGALE also gives values from the best-fitting model. For ID 258 in particular, while the SFR from Bayesian estimate is near SFMS, the SFR from best-fitting model is $0.01 \: M_\odot \: \mathrm{yr}^{-1}$, about 1 dex below the SFMS.

At the other end of the spectrum, the X-ray HAE ID 551 has a low SFR estimate but was not selected as a quiescent candidate based on its spectrum. This is due to its shallow $D_n4000$, which may be caused by the low signal-to-noise in the spectrum. However, the spectrum exhibits Balmer absorption lines with \OIII{} in emission, similar to ID 569, indicating that it might also be an AGN-hosting quenched galaxy.

These discrepancies demonstrate that additional tests are required to finally assess the true nature, e.g., by fitting spectroscopy and photometry data simultaneously, by constraining the level of obscured star formation of these galaxies, or/and by investigating spatially-resolved star-formation activity. Far-infrared and submillimeter data will help in clarifying the nature of these red galaxies. We defer deeper analyses to forthcoming papers. 

\subsection{Quiescent and AGN fractions}

We have found \numberquiescent{} quiescent galaxies identified from their grism spectra in the core of the protocluster. In $M_\star > 10^{11} M_\odot$ range, the fraction of quiescent galaxies is \quiescentfraction{} after excluding the three spectroscopically selected quiescent galaxies with high SFRs derived from SED fitting. This fraction comes down to \quiescentfractionlikely{} when including likely members. In $10^{10.5} < M_\star / M_\odot < 10^{11}$ stellar mass bin, the fraction is similarly $60^{+26}_{-30}\%$. These fractions show about three times enhancement compared to the general field quiescent fraction in the same stellar mass bin at $z \sim 2.2$ in COSMOS2020 \citep{weaver_cosmos2020_2023}, although their classification is based on $UVJ$ diagram. Since the field of view of our HST observation is relatively small, we cannot select comparison field galaxies in the same redshift slice ($2.10<z<2.21$) from our dataset to identify the quiescent galaxies in a uniform manner. We also checked objects within $2.30<z<3.00$, selecting 21 galaxies with good spectra, but we find all galaxies in the sample have stellar masses $M_\star<10^{10.5} \; M_\odot$ after SED fitting, which is to be expected since such massive galaxies are rare at this epoch. Thus, we cannot make a fair comparison with our protocluster sample regarding the quiescent fraction with such sample  

At $z\sim2$, several clusters have been found to host spectroscopically confirmed quiescent galaxies \citep{gobat_wfc3_2013, willis_spectroscopic_2020}. For comparison with other protoclusters, \citet{ito_cosmos2020_2023} found a threefold excess of $M_\star > 10^{11} M_\odot$ quiescent galaxies in a protocluster at $z=2.77$ in the COSMOS field. \citet{mcconachie_spectroscopic_2022} found a quiescent fraction of $\sim70\%$ in MAGAZ3NE J095924+022537 at $z=3.37$ at the same stellar mass range. At an even higher redshift, \citet{tanaka_proto-cluster_2023} spectroscopically confirmed quiescent galaxies in a protocluster at $z=4$, resulting in an estimated fraction of $30\%$. While the selection criteria might be different in each work, these protoclusters serve as evidence of accelerated formation and evolution of quiescent galaxies in such overdense environment.

Among the quiescent sample, we found that 4 of them are X-ray sources based on \citet{tozzi_700_2022}, indicating that they are AGN hosts. One galaxy may also be an obscured AGN based on its MEx ratio. The abundance of AGN in massive quiescent galaxies in this protocluster is in line with the findings of \citet{olsen_evidence_2013} and \citet{ito_cosmos2020_2022}, supporting the idea that AGN feedback plays an important role for quenching massive galaxies at $z > 1.5$. Furthermore, the X-ray-luminous fraction of \agnfraction{} among massive quiescent galaxies may be higher than that in field of $\sim20\%$ \citep{olsen_evidence_2013}. One might suggest that dense environment may cause the abundance of AGN in protocluster galaxies, e.g., through interactions and mergers, thus making protocluster galaxies more likely to quench. However, while there is an excess of AGN in the Spiderweb protocluster \citep{tozzi_700_2022, shimakawa_new_2024}, some protoclusters at similar redshifts lack such AGN excess \citep[e.g.,][]{macuga_fraction_2019}, muddling the conclusion.

\subsection{Size and morphology of protocluster members} \label{sec:size}


\begin{figure*}[t]
    \centering
    \includegraphics[width=\textwidth]{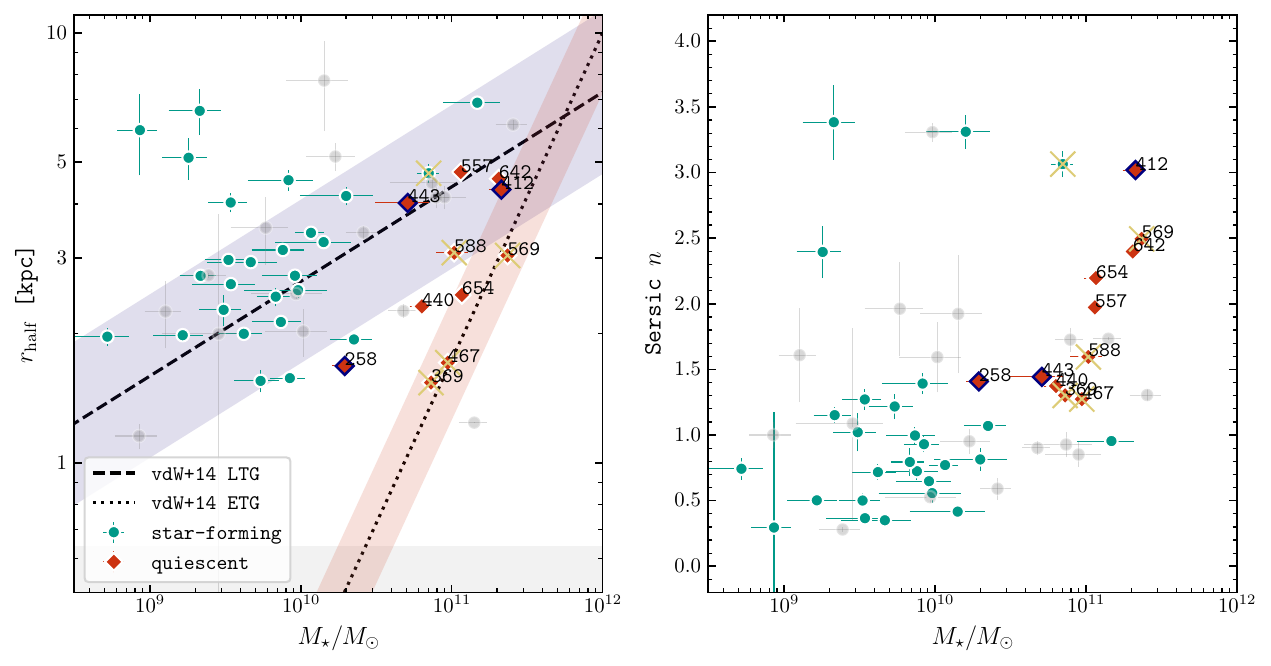}
    \caption{Sérsic half light radius (left) and Sérsic index (right) plotted against stellar mass. Mass-size relations from \citet{van_der_wel_3d-hstcandels_2014} are shown in dashed line (late-type galaxies) and dotted line (early-type galaxies), with shaded regions showing the intrinsic scatter. Red diamonds denotes spectroscopically selected quiescent galaxies, with blue borders indicating they are located near SFMS. Teal points are star-forming robust members and gray points are likely members. The quiescent galaxies are generally more compact than the star-forming mass-size relation. There seems to be a stellar mass--Sérsic index correlation, implying a regulation of light profile by stellar mass.}
    \label{fig:sersic}
\end{figure*}

We plot the half-light radius and Sérsic index against stellar mass in Figure \ref{fig:sersic}. Overall, the size of star-forming galaxies in the protocluster is consistent with the late-type galaxy stellar mass-size relation from \citet{van_der_wel_3d-hstcandels_2014} at $z = 2.2$. Quiescent members are generally more compact than the star-forming mass-size relation, with the exception of IDs 557 (MEx-AGN candidate) and 443 (possible dusty galaxy). These results are in agreement with the $K_s$-band size measurement by \citet{perez-martinez_signs_2023}, which found that red HAEs, which are X-ray sources, are more compact than regular HAEs at fixed stellar mass. We also note that the rest-frame optical size might be overestimated due to the presence of dust, as shown by \citet{suess_rest-frame_2022}.

On the right panel of Figure \ref{fig:sersic}, we show a plot of Sérsic index vs. stellar mass. The majority of non-quiescent members across stellar masses have disky morphology with $n \sim 1$. Quiescent members have higher Sérsic indices on average, but they are not as high as massive local ellipticals that generally exhibits index $n\gtrsim4$ \citep[e.g., ][]{lange_galaxy_2015}. Half of the quiescent galaxies have indices of $n\sim1.5$, indicating a significant disk component is present in each galaxy, while the other half may have started developing a bulge-like component. If we consider these quiescent galaxies as progenitors to massive local ellipticals, the rather intermediate Sérsic index values may indicate that quiescent galaxies are still in the morphological transition phase toward a more concentrated, cuspy profile, likely by dry minor mergers \citep{hilz_how_2013, newman_resolving_2018}.

While the difference of Sérsic index distribution of star-forming and quiescent galaxies might indicate the role of quenching in morphological transformation, we should note that there is a correlation between stellar mass and Sérsic index (Spearman $r_s=0.46$), and that there is a lack of robust, massive star-forming galaxies in our sample. Thus, it is possible that the light profile is dictated by stellar mass instead, not star-formation activity. If we include massive likely members to this analysis, 7 star-forming galaxies have $M_\star > 10^{10.5}$, compared to 8 quiescent galaxies in the same range. Assuming they are truly protocluster members and star-forming (see gray points in Figure \ref{fig:sfr}), Anderson-Darling test shows that the distribution of Sérsic index of quiescent and massive star-forming galaxies are unlikely to be drawn from the same parent distribution ($p=0.03$). This may hint at the role of star-formation quenching in the morphological transformation, but we reiterate the caveat of the inclusion of non-robust members in this analysis and the possibility of dust affecting the Sérsic profile measurement for massive star-forming galaxies. 

It should also be noted that progenitor bias may affect the comparison of size and morphology of star-forming galaxies and quiescent galaxies in the protocluster. Quiescent protocluster members may have formed earlier than the star-forming members, allowing them more time to develop bulges, as opposed to the result of star-formation quenching.

The activation of AGN has been linked to gas-rich major mergers \citep[e.g., ][see also \citealt{shah_investigating_2020, quai_interconnection_2023, koulouridis_agns_2024}]{ellison_galaxy_2013, satyapal_galaxy_2014, weston_incidence_2017}, which can lead to rapid quenching by AGN feedback. If quenching happens soon after merging, we might be able to detect merger signatures, as the observability timescale of mergers is $\sim 0.2-2$ Gyr depending on the initial condition \citep{lotz_galaxy_2008}, similar to the estimated age from $D_n4000$ of the quiescent galaxies in our sample. In low redshift Universe, \citet{ellison_galaxy_2022} find a significant excess of rapidly quenched galaxies among post-mergers based on ground-based Canada France Imaging Survey (CFIS) data. In higher redshifts, however, \citet{shah_investigating_2020} did not find significant enhancement of AGN activity in interacting galaxies.

Based on visual inspection of the morphologies of quiescent galaxies in our sample, we do not see a sign of major mergers such as tidal tails, with possible exceptions of ID 588 and ID 412, which exhibit a disturbance in their morphologies, and ID 369, which has a likely member as close neighbor ID 362 (see Figure \ref{fig:showcase1}). As we discussed in Section \ref{sec:quiescent}, ID 412 in particular is a part of a very dense system which will eventually coalesce into one massive galaxy. It is possible that it is merging with another member of the system, which results in the disturbance in the apparent morphology. IDs 588 and 369 are both AGN hosts, which may be triggered by interactions with their close neighbors \citep[e.g., ][]{ellison_galaxy_2013, weston_incidence_2017}. In fact, 5 out of 6 robust members which are X-ray sources have a close neighboring galaxy within $2 \text{ arcsec}$ ($\approx 17 \text{ kpc}$ at $z\sim2.16$), although in most cases they are fainter galaxies with $P_\mathrm{cl} < 0.5$. In contrast, only 12 out of 34 non-X-ray robust members have neighbors within such separation. This might be  evidence of the link between galaxy interactions, AGN activity, and quenching, which may be induced by the overdense environment in the protocluster.

Quantitative morphology statistics such as the Concentration-Asymmetry-Smoothness system \citep{conselice_asymmetry_2000} and Gini-$M_{20}$ \citep{lotz_new_2004} using \textsc{Statmorph} \citep{rodriguez-gomez_optical_2019} have also been used in the literature to identify disturbances in the morphologies as a proxy of merger signatures \citep{conselice_evolution_2014, peth_beyond_2016, sazonova_morphology-density_2020, naufal_environmental_2023, laishram_insights_2024}. In particular, \citet{naufal_environmental_2023} finds an evidence of higher disturbance in protocluster galaxies. While their sample includes galaxies in the Spiderweb protocluster, their sample consists only of HAEs and the morphologies were measured in rest-frame UV. JWST NIRCam data will be needed to provide the rest-frame near-infrared view to assess the true stellar distribution in galaxies in the protocluster.


\section{Summary} \label{sec:conclusion}
We report the results of a survey of Spiderweb protocluster core with deep HST WFC3 G141 slitless spectroscopy observation. Based on grism redshift determination with \textsc{Grizli}, we identified \robustmembers{} galaxies as robust members of the protocluster with $H_{160} < 25$, 19 of which are previously identified as HAE members from narrowband selection by \citet{shimakawa_mahalo_2018-1}. We also spectroscopically identified new \OIII-emitters as members of the protocluster and new quiescent galaxies selected by the strength of $4000 \text{ \AA}$ break in their spectra. We confirmed the overdensity of galaxies previously found from HAEs, although we do not find an overdensity in the extended redshift range $2.10 < z < 2.21$ as suggested by \citet{jin_coalas_2021} based on CO-emitters due to the smaller field of view. 

The observation reveals \numberquiescent{} galaxies with quiescent spectra in the protocluster core. Three of these galaxies may still be star-forming according to SED fitting despite their quiescent spectra. These may result from leftover star-formation after a starbursting phase, or from a dusty galaxy mimicking as quiescent one.

We estimate the fraction of quiescent galaxies in the core of the protocluster to be $\sim60\%$ for $M_\star \geq 10^{11}$, about three times higher than that in the general field. These quiescent protocluster members exhibit somewhat more compact sizes and more concentrated light profiles than than the star-forming members, but are still not as cuspy as quiescent elliptical galaxies in the local Universe. This may indicate that these quiescent members have not yet experience enough transformation, e.g., from dry mergers, to make them structurally similar to local elliptical galaxies.

AGN feedback has been popularly thought to play an important role in quenching massive galaxies at high redshift. Half of the quiescent galaxies in the Spiderweb protocluster are indicated to host AGN. Such high fraction of AGN among massive quiescent galaxies may also be higher than that in field \citep{olsen_evidence_2013}, which may be induced by the overdense environment.

\facilities{Hubble Space Telescope, Subaru Telescope}

\software{\textsc{Astropy} \citep{astropy_collaboration_astropy_2013, astropy_collaboration_astropy_2018, astropy_collaboration_astropy_2022}, \textsc{Grizli} \citep{brammer_grizli_2019}, \textsc{CIGALE} \citep{boquien_cigale_2019}, \textsc{GALFIT} \citep{peng_detailed_2010}, \textsc{SExtractor} \citep{bertin_sextractor_1996}, \textsc{Matplotlib} \citep{hunter_matplotlib_2007}}

\begin{acknowledgements}
We thank the anonymous referee for their constructive feedback. CDE thanks F. Rizzo for helpful discussions. This research is based on observations made with the NASA/ESA Hubble Space Telescope obtained from the Space Telescope Science Institute, which is operated by the Association of Universities for Research in Astronomy, Inc., under NASA contract NAS 5–26555. These observations are associated with program ID17117, PI: Yusei Koyama. Some/all of the data presented in this article were obtained from the Mikulski Archive for Space Telescopes (MAST) at the Space Telescope Science Institute. The specific observations analyzed can be accessed via \dataset[doi: 10.17909/2kxb-5r67]{https://doi.org/10.17909/2kxb-5r67}. This research is based [in part] on data collected at the Subaru Telescope, which is operated by the National Astronomical Observatory of Japan. We are honored and grateful for the opportunity of observing the Universe from Maunakea, which has the cultural, historical, and natural significance in Hawaii. 

This work was supported by JSPS KAKENHI Grant Number J23H01219 and JSPS Core-to-Core Program (grant number: JPJSCCA20210003). HD, JMPM and YZ acknowledges financial support from the Agencia Estatal de Investigación del Ministerio de Ciencia e Innovación (AEI-MCINN) under grant (La evolución de los cúmulos de galaxias desde el amanecer hasta el mediodía cósmico) with reference (PID2019-105776GB-I00/DOI:10.13039/501100011033) and Agencia Estatal de Investigación del Ministerio de Ciencia, Innovación y Universidades (MCIU/AEI) under grant (Construcción de cúmulos de galaxias en formación a través de la formación estelar oscurecida por el polvo) and the European Regional Development Fund (ERDF) with reference (PID2022-143243NB-I00/10.13039/501100011033). CDE acknowledges funding from the MCIN/AEI (Spain) and the “NextGenerationEU”/PRTR (European Union) through the Juan de la Cierva-Formación program (FJC2021-047307-I). JMPM acknowledges funding from the European Union’s Horizon-Europe research and innovation programme under the Marie Skłodowska-Curie grant agreement No 101106626. YZ acknowledges the support from the China Scholarship Council (202206340048), and the National Science Foundation of Jiangsu Province (BK20231106).

\end{acknowledgements}

\bibliography{references}{}
\bibliographystyle{aasjournal}



\appendix

\section{Likely members} \label{sec:likely}

\begin{figure}
    \centering
    \includegraphics[width=0.5\linewidth]{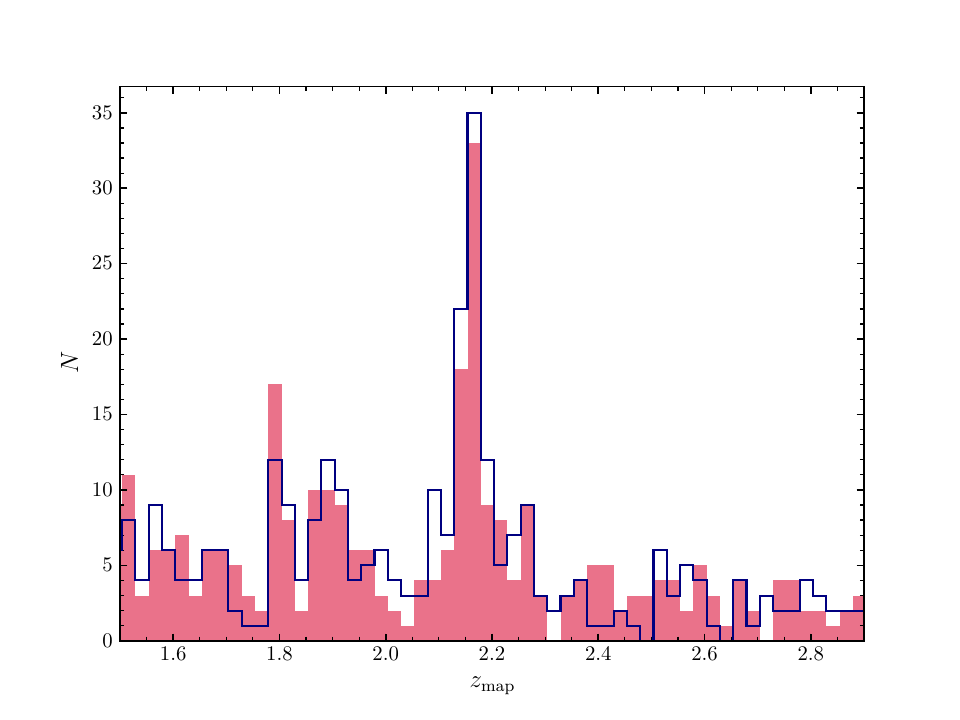}
    \caption{Redshift histograms of the default source detection (gray) and the alternative source detection (red) parameters. The different parameters lead to different redshift determination, and we consider `robust' members as objects whose redshifts from the two configurations are within $2.10 < z < 2.20$.}
    \label{fig:alt_redshift_hist}
\end{figure}

As we mentioned in Section \ref{sec:membership}, we ran the pipeline with alternate source detection parameters. 
In default settings, \textsc{Grizli} runs the detection using \textsc{deblend\_cont} $= 0.001$. We find that using a more lenient deblending of \textsc{deblend\_cont} $=0.01$ leads to different redshift determination by \textsc{Grizli}. We attribute this to different contamination models produced by the two configurations, which affects clean spectrum extraction and eventually the redshift determination (see Figure \ref{fig:alt_redshift_hist}).

Objects considered as a member only by one configuration of source detection is considered as a `likely' member. In addition, HAEs identified by \citetalias{shimakawa_mahalo_2018-1} that are not recovered by \textsc{Grizli} are also considered as `likely' members if their spectra show typical features for galaxies at $z\sim2.16$ such as redshifted \OIII-like emission line (i.e., an emission line at $\lambda \sim 15800$). We show the list of likely members in Table \ref{tab:catalog_likely} along with some comments on each object.

\begin{deluxetable*}{cccccl}
\tablecaption{Galaxies considered as likely members of Spiderweb protocluster.} \label{tab:catalog_likely}
\tablehead{\colhead{ID} &  \colhead{$z_\mathrm{grism}$} &  \colhead{$H_{160}$} & \colhead{ID \citetalias{shimakawa_mahalo_2018-1}} &  \colhead{class} &   \colhead{comments}}
\startdata
266 &   2.191 & 24.485 &    -- &    likely &    Bad fit quality: FSPS template fits worse than a smooth polynomial template. \\
314 &   1.685 & 23.906 &    22 &    likely &    HAE. The alternate configuration shows a quiescent-like spectrum at $z\approx 2.3$.\\
341 &   0.567 & 24.050 &    -- &    likely &    Strong \OIII-like emission. Identified as a member by alternate configuration at $z=2.147$. \\
362 & $2.189$ & 23.093 &    28 &    likely &    HAE. $P_\mathrm{cl} < 0.5$. Quiescent spectrum; strong $4000 \text{ \AA}$ break and possible Balmer absorption lines. \\
471 &   4.633 & 23.824 &    36 &    likely &    HAE. Strong \OIII-like emission, but the $11000 - 12000 \text{ \AA}$ region is contaminated. \\
505 &   1.341 & 20.943 &    46 &    likely &    HAE. Bright blue continuum prevents \textsc{Grizli} to identify it as a member despite Hb+OIII lines. \\
526 &   0.567 & 24.491 &    -- &    likely &    Strong \OIII-like line. Identified as a member by alternate configuration at $z=2.14$. \\
541 &   2.295 & 22.916 &    -- &    likely &    Possibly a dusty galaxy. Considered as one object by alternate configuration at $z=2.188$. \\
542 &   2.152 & 23.135 &    -- &    likely &    Possibly a dusty galaxy. Considered as one object by alternate configuration at $z=2.188$. \\
545 &   2.218 & 21.851 &    54 &    likely &    HAE. A dusty galaxy. \\
595 &   2.190 & 24.156 &    56 &    likely &    HAE. $P_\mathrm{cl} < 0.5$ and $H_{160} > 24$. Not selected as a member by alternate configuration. \\
624 &   2.149 & 24.473 &    -- &    likely &    Identified as a member in main pipeline only. \\
709 &   0.233 & 23.873 &    67 &    likely &    HAE. Shallow spectrum, but it has a strong OIII emission. \\
773 &   0.548 & 24.778 &    -- &    likely &    Strong \OIII-like emission. Identified as a member by alternate configuration at $z=2.108$. \\
\enddata

\end{deluxetable*}

\end{document}